\documentclass[twocolumn,showpacs,preprintnumbers,superscriptaddress,floatfix,nofootinbib]{revtex4}
\usepackage{amssymb}
\usepackage{amsmath}
\usepackage{graphicx}
\usepackage{dcolumn}
\usepackage{bm}
\usepackage[subfigure]{graphfig}
\usepackage{makecell}

\begin{document}

\title{Production of the superheavy baryon $\Lambda _{c\bar{c}}^{\ast
}(4209) $ in kaon-induced reaction}
\author{Xiao-Yun Wang}
\thanks{Corresponding author: xywang@impcas.ac.cn}
\affiliation{Institute of Modern Physics, Chinese Academy of Sciences, Lanzhou 730000,
China}
\affiliation{University of Chinese Academy of Sciences, Beijing 100049, China}
\affiliation{Research Center for Hadron and CSR Physics, Institute of Modern Physics of
CAS and Lanzhou University, Lanzhou 730000, China}
\author{Xu-Rong Chen}
\affiliation{Institute of Modern Physics, Chinese Academy of Sciences, Lanzhou 730000,
China}
\affiliation{Research Center for Hadron and CSR Physics, Institute of Modern Physics of
CAS and Lanzhou University, Lanzhou 730000, China}

\begin{abstract}
The production of superheavy $\Lambda _{c\bar{c}}^{\ast }(4209)$ baryon in
the $K^{-}p\rightarrow \eta _{c}\Lambda $ process via $s$-channel is
investigated with an effective Lagrangian approach and the isobar model.
Moreover, the background from the $K^{-}p\rightarrow \eta _{c}\Lambda $
reaction through the $t$-channel with $K^{\ast }$ exchange and $u$-channel
with nucleon exchange are also considered. The numerical results indicate it
is feasible to search for the superheavy $\Lambda _{c\bar{c}}^{\ast }(4209)$
via $K^{-}p$ scattering. The relevant calculations not only shed light on
the further experiment of searching for the $\Lambda _{c\bar{c}}^{\ast
}(4209)$ through kaon-induced reaction, but enable us to have a better
understanding of the exotic baryons.
\end{abstract}

\maketitle

\section{Introduction}

At present, searching and explaining the exotic states have becoming a very
interesting issue in hadron physics. These exotic states are cannot be
included in the conventional picture of $qqq$ for baryons and $q\bar{q}$ for
mesons but allowed to exist in the particle family when taking into account
the color and flavor degrees of freedom in quark models.

With the development of experiments, a series of charmoniumlike and
bottomoniumlike mesons (which were named as XYZ states) have been observed
\cite{lx14,nb11}. Since these XYZ mesons definitely cannot be accommodated
into the frame of conventional charmonium $c\bar{c}$ states, they are
considered as the most promising candidate of the exotic states.
Theoretically, abundant investigation were carried out on the nature and
production mechanism of XYZ states \cite{lx14,nb11,xyw15,xy}. They are
interpreted as a tetraquark, a hadron molecule or just a cusp effect, etc
\cite{lx14,nb11}.

In addition, the scientific study of multiquark baryons was already in full
swing. In the early 1980s, Brodsky \textit{et al. }proposed that there exist
a few non-negligible intrinsic $uudc\bar{c}$ components ($\sim 1\%$) in the
proton \cite{bro80}. Later, a new measurement about parity-violating
electron scattering (PVES) at JLab provides a direct evidence for the
existence of the multiquark components in the proton \cite{aa07}. Moreover,
in order to explain the phenomenon that the strong coupling of $N^{\ast
}(1535)$ to the final states with strangeness and the mass order between $%
N^{\ast }(1535)$ and $\Lambda ^{\ast }(1405)$ \cite{pdg,bs09}, the
pentaquark picture was proposed by theorists\cite{bs09,ch02,bs10}, which
makes that the hidden strange $N^{\ast }(1535)$ with $\left\vert uuds\bar{s}%
\right\rangle $ is naturally heavier than the open strange $\Lambda ^{\ast
}(1405)$ with $\left\vert udsq\bar{q}\right\rangle $. However, due to
tunable ingredients and possible large mixing of various configurations,
none of them can be clearly distinguished from the conventional $qqq$ or $q%
\bar{q}$ states. A hopeful solution for this problem is to extend pentaquark
to hidden charm and hidden beauty for baryons.

In refs. \cite{jj10,jj11,jj12}, within the framework of the coupled-channel
unitary approach with the local hidden gauge formalism, several narrow
hidden charm $N_{c\bar{c}}^{\ast }$ and $\Lambda _{c\bar{c}}^{\ast }$
resonances were predicted with mass above 4 GeV and width smaller than 100
MeV. Soon after, some hidden beauty $N_{b\bar{b}}^{\ast }$ and $\Lambda _{b%
\bar{b}}^{\ast }$ baryons were also proposed in ref. \cite{jjplb12}. These
predicted baryons are dynamically generated in the $PB$ and $VB$ channels,
where $P$ and $V$ stand for the pseudoscalar and vector mesons, respectively
\cite{jj10,jj11,jj12,jjplb12}. In these predicted baryons, the production of
$N^{\ast }(4261)$ resonance by $\pi ^{-}p$ collision has been studied in our
previous work \cite{epl15}. Besides, one find that the hidden charm $\Lambda
_{c\bar{c}}^{\ast }(4209)$ and hidden beauty $\Lambda _{b\bar{b}}^{\ast
}(11021)$ baryons all have a large coupling with $\bar{K}N$ (the detailed
informations about these two resonances are listed in the table I), which
mean that searching for these superheavy baryons via $K^{-}p$ scattering
will be feasible and effective. Since the a maximum $K$ beam of 20 GeV$/c$
can be produced at J-PARC, it allows one to observe the $\Lambda _{c\bar{c}%
}^{\ast }(4209)$ but not for the $\Lambda _{b\bar{b}}^{\ast }(11021)$
states. Thus we only focus on the production of $\Lambda _{c\bar{c}}^{\ast
}(4209)$ in the present work.

\begin{table*}[tbph]
\caption{Informations of the $\Lambda _{c\bar{c}}^{\ast }(4209)$ and $%
\Lambda _{b\bar{b}}^{\ast }(11021)$ states from $PB\rightarrow PB$ as
predicted in refs. \protect\cite{jj10,jj11,jj12,jjplb12}. $M$, $\Gamma $ and
$\Gamma _{i}$ stand for the mass, total width, and partial decay width,
respectively. }%
\begin{tabular}{p{2cm}p{2cm}p{2cm}p{1cm}p{1cm}p{1cm}p{1cm}p{1cm}p{1cm}p{1cm}p{1cm}}
\hline\hline
& $M$(MeV) & $\Gamma $(MeV) &  &  &  & $\Gamma _{i}$(MeV) &  &  &  &  \\
\hline
&  &  & $\bar{K}N$ & $\pi \Sigma $ & $\eta \Lambda $ & $\eta ^{\prime
}\Lambda $ & $K\Xi $ & $\eta _{c}\Lambda $ & $\eta _{b}\Lambda $ & $%
B_{s}\Lambda _{b}$ \\
$\Lambda _{c\bar{c}}^{\ast }(4209)$ & 4209 & 32.4 & 15.8 & 2.9 & 3.2 & 1.7 &
2.4 & 5.8 & --- & --- \\
$\Lambda _{b\bar{b}}^{\ast }(11021)$ & 11021 & 2.21 & 0.65 & 0.01 & 0.08 &
0.14 & 0.01 & --- & 0.19 & 1.18 \\ \hline\hline
\end{tabular}%
\end{table*}

In this paper, with an effective Lagrangian approach and isobar model, we
study the role and production of $\Lambda _{c\bar{c}}^{\ast }(4209)$ in the $%
K^{-}p\rightarrow \eta _{c}\Lambda $ process. Moreover, the feasibility of
searching for the superheavy $\Lambda _{c\bar{c}}^{\ast }(4209)$ resonance
is also discussed. It is shown that the J-PARC \cite{jparc} will be an ideal
platform for searching for the superheavy $\Lambda _{c\bar{c}}^{\ast }(4209)$
baryon, which will hopefully confirm our numerical predictions for $\Lambda
_{c\bar{c}}^{\ast }(4209)$ production.

This paper is organized as follows. After an Introduction, we will present
the formalism and the main ingredients which are used in our calculation.
The numerical results and discussions are given in Sec. III. Finally, the
paper ends with a brief summary.

\section{Formalism}

In this work, an effective Lagrangian approach and isobar model in terms of
hadrons are used in our calculation, which is an important theoretical
method in investigating various processes in the resonance region. Fig. 1(a)
describes the basic tree level Feynman diagrams for the production of $%
\Lambda _{c\bar{c}}^{\ast }(4209)$ in $K^{-}p\rightarrow \eta _{c}\Lambda $
reaction through $s$-channel. The background contributions are mainly from
the $t$-channel $K^{\ast }$ meson exchange and $u$-channel proton exchange,
as depicted in Fig. 1(b).

\begin{figure}[tbph]
\begin{center}
\includegraphics[scale=0.5]{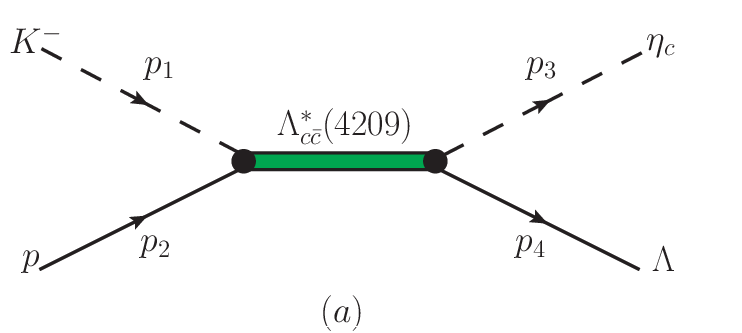} \includegraphics[scale=0.5]{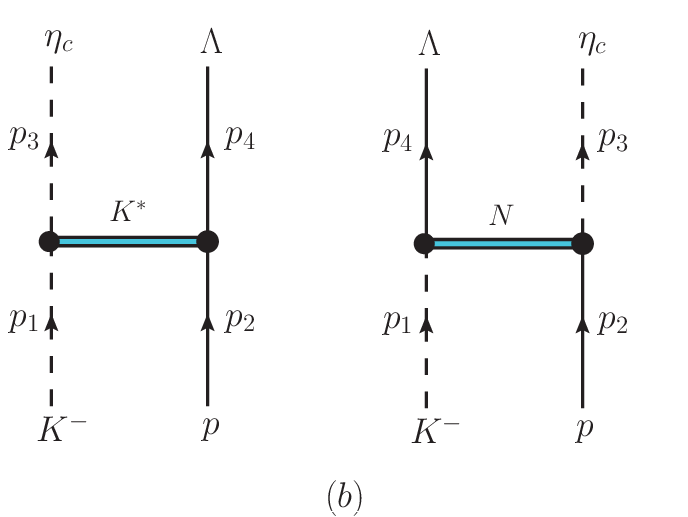}
\end{center}
\caption{(Color online) Feynman diagrams for the $K^{-}p\rightarrow \protect%
\eta _{c}\Lambda $ reaction.}
\label{Fig:fyd}
\end{figure}
Since the spin-parity ($J^{P}$) quantum number of $\Lambda _{c\bar{c}}^{\ast
}(4209)$ was determined to be $\frac{1}{2}^{-}$ \cite{jj10,jj11,jj12}, for
the intermediate $\Lambda _{c\bar{c}}^{\ast }(4209)$ ($\Lambda ^{\ast }$ for
short) resonance contribution in the $s$ channel, we take the normally used
effective Lagrangians for $KN\Lambda ^{\ast }$ and $\eta _{c}\Lambda \Lambda
^{\ast }$ couplings as \cite{zou03,bc12},

\begin{eqnarray}
\mathcal{L}_{KN\Lambda ^{\ast }} &=&g_{_{KN\Lambda ^{\ast }}}\bar{K}\bar{%
\Lambda}^{\ast }N+h.c., \\
\mathcal{L}_{\eta _{c}\Lambda \Lambda ^{\ast }} &=&g_{\eta _{c}\Lambda
\Lambda ^{\ast }}\bar{\Lambda}^{\ast }\Lambda \eta _{c}+h.c.,
\end{eqnarray}%
The coupling constant $g_{_{KN\Lambda ^{\ast }}}$ and $g_{_{\eta _{c}\Lambda
\Lambda ^{\ast }}}$ are determined by the partial decay widths of $\Lambda
_{c\bar{c}}^{\ast }(4209),$%
\begin{eqnarray}
\Gamma _{\Lambda ^{\ast }\rightarrow NK} &=&\frac{g_{_{KN\Lambda ^{\ast
}}}^{2}(m_{N}+E_{N})}{4\pi M_{\Lambda ^{\ast }}}\left\vert \vec{p}_{N}^{~%
\mathrm{c.m.}}\right\vert , \\
\Gamma _{\Lambda ^{\ast }\rightarrow \Lambda \eta _{c}} &=&\frac{g_{_{\eta
_{c}\Lambda \Lambda ^{\ast }}}^{2}(m_{\Lambda }+E_{\Lambda })}{4\pi
M_{\Lambda ^{\ast }}}\left\vert \vec{p}_{\Lambda }^{~\mathrm{c.m.}%
}\right\vert ,
\end{eqnarray}%
with%
\begin{eqnarray}
|\vec{p}_{N}^{~\mathrm{c.m.}}| &=&\frac{\lambda ^{1/2}(M_{\Lambda ^{\ast
}}^{2},m_{N}^{2},m_{K}^{2})}{2M_{\Lambda ^{\ast }}}, \\
E_{N} &=&\sqrt{|\vec{p}_{N}^{~\mathrm{c.m.}}|^{2}+m_{N}^{2}}, \\
|\vec{p}_{\Lambda }^{~\mathrm{c.m.}}| &=&\frac{\lambda ^{1/2}(M_{\Lambda
^{\ast }}^{2},m_{\Lambda }^{2},m_{\eta _{c}}^{2})}{2M_{\Lambda ^{\ast }}}, \\
E_{\Lambda } &=&\sqrt{|\vec{p}_{\Lambda }^{~\mathrm{c.m.}}|^{2}+m_{\Lambda
}^{2}},
\end{eqnarray}%
where $\lambda $ is the K$\ddot{a}$llen function with $\lambda
(x,y,z)=(x-y-z)^{2}-4yz$. Using the predicted partial decay widths of $%
\Lambda _{c\bar{c}}^{\ast }(4209)$ \cite{jj10,jj11} as listed in Table I,
one gets $g_{_{KN\Lambda ^{\ast }}}=0.369$ GeV$^{-1}$ and $g_{_{\eta
_{c}\Lambda \Lambda ^{\ast }}}=0.557$ GeV$^{-1}$.

For the $t$-channel $K^{\ast }$ meson exchange, the effective Lagrangian for
the $K^{\ast }K\eta _{c}$ coupling is%
\begin{equation}
\mathcal{L}_{K^{\ast }K\eta _{c}}=ig_{_{K^{\ast }K\eta _{c}}}\bar{K}_{\mu
}^{\ast }(\eta _{c}\partial ^{\mu }\bar{K}-\bar{K}\partial ^{\mu }\eta
_{c})+h.c.,
\end{equation}%
where the isospin structure for $K^{\ast }K\eta _{c}$ is $\bar{K}^{\ast }%
\bar{K}\eta _{c}$ with%
\begin{equation}
\bar{K}^{\ast }=(K^{\ast -},\bar{K}^{\ast 0}),\text{ \ }\bar{K}=\binom{K^{-}%
}{\bar{K}^{0}}.
\end{equation}%
With the partial decay width $\Gamma _{\eta _{c}\rightarrow K^{\ast }\bar{K}%
}<32.2\times 1.28\%$ (MeV)\cite{pdg}, one obtains the upper limit of
constant coupling: $g_{_{K^{\ast }K\eta _{c}}}=0.203$ GeV$^{-1}$.

The effective Lagrangian for $K^{\ast }N\Lambda $ vertex is%
\begin{equation}
\mathcal{L}_{K^{\ast }N\Lambda }=-g_{K^{\ast }N\Lambda }\bar{\Lambda}(\gamma
_{\mu }K^{\ast \mu }-\frac{\kappa _{K^{\ast }N\Lambda }}{2m_{N}}\sigma _{\mu
\nu }\partial ^{\nu }K^{\ast \mu })N+h.c.,
\end{equation}%
where the values of $g_{K^{\ast }N\Lambda }$ and $\kappa _{K^{\ast }N\Lambda
}$ were given in many theoretical works. In refs.\cite{vg99,yo06}, two sets
of these coupling constants are obtained by the potential model, namely,%
\begin{eqnarray}
g_{K^{\ast }N\Lambda } &=&-4.26,\text{ \ \ }\kappa _{K^{\ast }N\Lambda
}=2.66,  \notag \\
g_{K^{\ast }N\Lambda } &=&-6.11,\text{ \ \ }\kappa _{K^{\ast }N\Lambda
}=2.43.
\end{eqnarray}%
Besides, one notice that the experiment datas for the $K^{-}p\rightarrow \pi
^{0}\Lambda $ process were produced very well by taking $g_{K^{\ast
}N\Lambda }=-6.11$ and $g_{K^{\ast }N\Lambda }\cdot \kappa _{K^{\ast
}N\Lambda }=-11.33$ in ref. \cite{pu11}. In this work, we take $g_{K^{\ast
}N\Lambda }=-6.11$ and $\kappa _{K^{\ast }N\Lambda }=1.85$ in order to
obtain a more accurate estimates.

For the $u$-channel nucleon exchange, the effective Lagrangians for $%
KN\Lambda $ and $\eta _{c}NN$ couplings are%
\begin{eqnarray}
\mathcal{L}_{KN\Lambda } &=&\frac{g_{KN\Lambda }}{m_{N}+m_{\Lambda }}\bar{N}%
\gamma ^{\mu }\gamma _{5}\Lambda \partial _{\mu }K+h.c., \\
\mathcal{L}_{\eta _{c}NN} &=&g_{_{\eta _{c}NN}}\bar{N}\gamma _{5}\eta
_{c}N+h.c.,
\end{eqnarray}%
where $g_{KN\Lambda }=-13.24$ are estimated from flavor SU(3) symmetry
relations \cite{pu11,oh06,oh08}. The coupling constant $g_{_{\eta _{c}NN}}$
is determined from the partial decay widths of $\eta _{c},$%
\begin{equation}
\Gamma _{\eta _{c}\rightarrow p\bar{p}}=\left( g_{_{\eta _{c}NN}}\right) ^{2}%
\frac{\left\vert \vec{p}^{~\mathrm{c.m.}}\right\vert }{8\pi },
\end{equation}%
with%
\begin{equation}
\left\vert \vec{p}^{~\mathrm{c.m.}}\right\vert =\frac{\sqrt{M_{\eta
_{c}}^{2}-4M_{p}^{2}}}{2}.
\end{equation}%
where $\left\vert \vec{p}^{~\mathrm{c.m.}}\right\vert $ is three momentum of
nucleon in the rest frame of $\eta _{c}$ meson. With $\Gamma _{\eta
_{c}\rightarrow p\bar{p}}=49.1$ keV~\cite{pdg}, one obtain $g_{_{\eta
_{c}NN}}=3.26\times 10^{-2}$ GeV$^{-1}$.

Considering the internal structure of hadrons, a form factor is introduced
to describe the possible off-shell effects in the amplitudes. For $s$ and $u$
channels, we adopt the following form factors as used in refs. \cite%
{bc12,pu11,mosel98,mosel99,vs05},%
\begin{equation}
\mathcal{F}_{s/u}(q_{ex}^{2})=\frac{\Lambda _{s/u}^{4}}{\Lambda
_{s/u}^{4}+(q_{ex}^{2}-M_{ex}^{2})^{2}},
\end{equation}%
while for the $t$-channel, we take

\begin{equation}
\mathcal{F}_{t}(q_{ex}^{2})=\frac{\Lambda _{t}^{2}-M_{ex}^{2}}{\Lambda
_{t}^{2}-q_{ex}^{2}}
\end{equation}%
where $q_{ex}$ and $M_{ex}$ are the four-momenta and the mass of the
exchanged hadron, respectively. The values of cutoff parameters $\Lambda $ $%
(\Lambda _{s},\Lambda _{u}$ and $\Lambda _{t})$ will be discussed in the
next section.

For the propagators with four-momenta $q_{ex},$ we adopt the Breit-Wigner
form \cite{bc12,wh02}%
\begin{equation}
G_{\Lambda ^{\ast }}(q_{ex})=i\frac{\rlap{$\slash$}q_{ex}+M_{\Lambda ^{\ast
}}}{q_{ex}^{2}-M_{\Lambda ^{\ast }}^{2}+iM_{\Lambda ^{\ast }}\Gamma
_{\Lambda ^{\ast }}}
\end{equation}
for the superheavy baryon $\Lambda _{c\bar{c}}^{\ast }(4209)$, where $\Gamma
_{\Lambda ^{\ast }}$ is the total decay width of $\Lambda _{c\bar{c}}^{\ast
}(4209)$ state.

For the $K^{\ast }$ in the $t$-channel, we take the propagator as
\begin{equation}
G_{K^{\ast }}(q_{ex})=i\frac{-g^{\mu \nu }+q_{ex}^{\mu }q_{ex}^{\nu
}/M_{K^{\ast }}^{2}}{q_{ex}^{2}-M_{K^{\ast }}^{2}},
\end{equation}%
where $\mu $ and $\nu $ denote the polarization indices of vector meson $%
K^{\ast }.$

For the nucleon propagator, we take%
\begin{equation}
G_{N}(q_{ex}^{2})=i\frac{\rlap{$\slash$}q_{ex}+M_{N}}{q_{ex}^{2}-M_{N}^{2}}
\end{equation}

With the Feynman rules, the invariant scattering amplitude for the $%
K^{-}(p_{1})p(p_{2})\rightarrow \eta _{c}(p_{3})\Lambda (p_{4})$ reaction as
shown in Fig. 1 can be constructed as,%
\begin{eqnarray}
\mathcal{M}_{i} &\mathcal{=}&\overline{u}_{r_{2}}(p_{4})\mathcal{A}%
u_{r_{1}}(p_{2})  \notag \\
&=&\overline{u}_{r_{2}}(p_{4})\left( \sum\limits_{i}\mathcal{A}_{i}\right)
u_{r_{1}}(p_{2}),
\end{eqnarray}%
where $i$ denotes the $s$-, $t$-, or $u$-channel process that contribute to
the total amplitude, while $\overline{u}_{r_{2}}(p_{4})$ and $%
u_{r_{1}}(p_{2})$ are the spinors of the outgoing $\Lambda $ baryon and the
initial proton, respectively.

We define $s=(p_{1}+p_{2})^{2}\equiv W^{2}$, then the unpolarized
differential cross section for the $K^{-}p\rightarrow \eta _{c}\Lambda $
reaction at the center of mass (c.m.) frame is given by
\begin{equation}
\frac{d\sigma }{d\cos \theta }=\frac{1}{32\pi s}\frac{\left\vert \vec{p}%
_{3}^{~\mathrm{c.m.}}\right\vert }{\left\vert \vec{p}_{1}^{{~\mathrm{c.m.}}%
}\right\vert }\left( \frac{1}{2}\sum\limits_{r_{1},r_{2}}\left\vert \mathcal{%
M}\right\vert ^{2}\right)
\end{equation}%
where $\theta $ denotes the angle of the outgoing $\eta _{c}$ meson relative
to beam direction in the c.m. frame, while $\vec{p}_{1}^{~\mathrm{c.m.}}$
and $\vec{p}_{3}^{~\mathrm{c.m.}}$ are the three-momenta of initial $K^{-}$
and final $\eta _{c}$, respectively. Therefore, the total invariant
scattering amplitude $\mathcal{M}$ is written as,%
\begin{equation}
\mathcal{M=M}_{s}^{\Lambda ^{\ast }}+\mathcal{M}_{t}^{K^{\ast }}+\mathcal{M}%
_{u}^{N}.
\end{equation}%
It is important to note that in phenomenological approaches, the relative
phases between different amplitudes are not fixed. Since we do not now have
experimental data, and we will see in the following that the magnitudes of
signal and background contributions in the energy region that we considered
are much different, the effect of the interference term should be small.
Thus we take all the relative phases as zero in the present work.

\section{Results and discussion}

With the formalism and ingredients given above, the total cross sections for
the $K^{-}p\rightarrow \eta _{c}\Lambda $ reaction are calculated. Since the
cutoff parameter $\Lambda $ related to the form factor is the only free
parameter, we first need to give a constraint on the value of $\Lambda $.
\begin{figure}[tbph]
\centering
\includegraphics[scale=0.4]{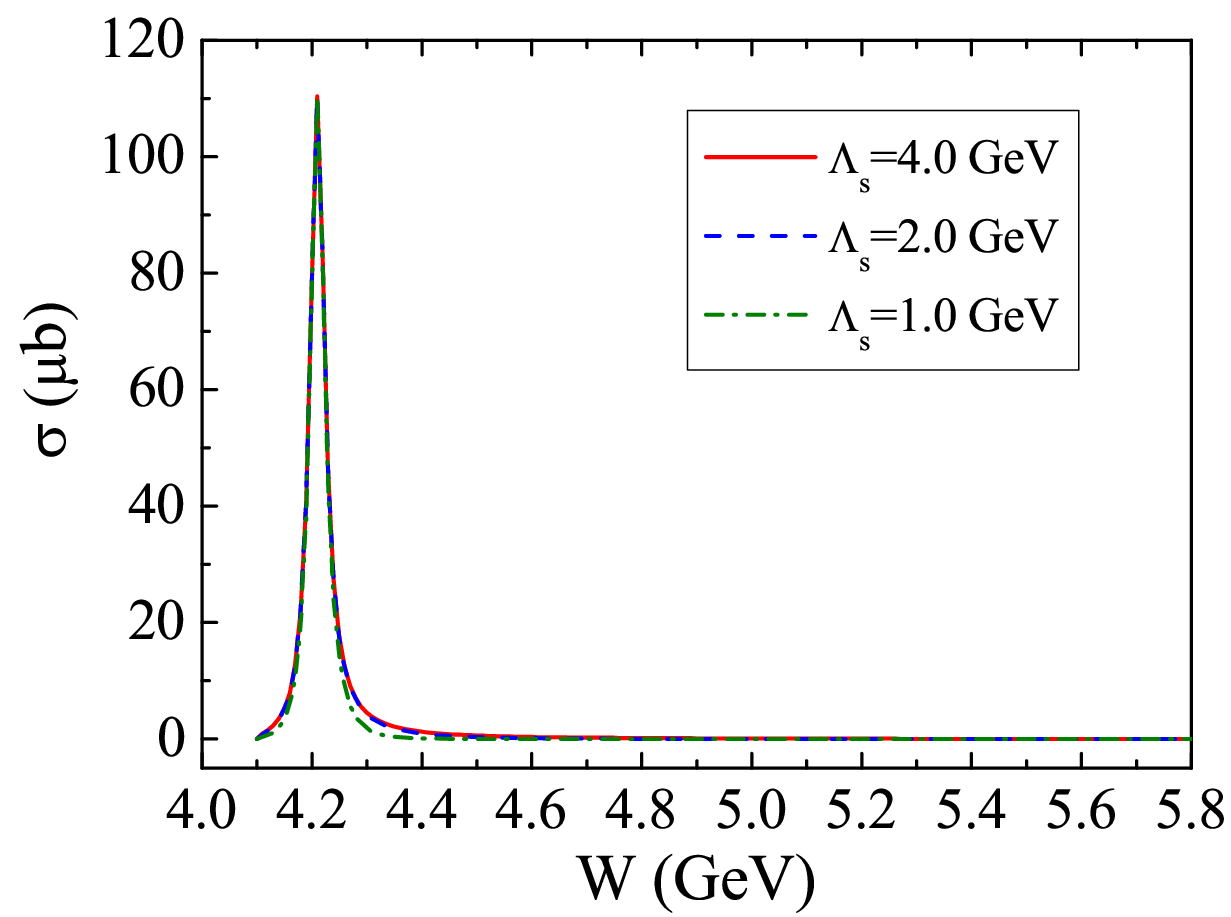}
\caption{(Color online) The cross section for the production of $\Lambda _{c%
\bar{c}}^{\ast }(4209)$ through s-channel with the different typical cut-off
$\Lambda _{s}$.}
\end{figure}

\begin{figure}[tbph]
\begin{center}
\includegraphics[scale=0.35]{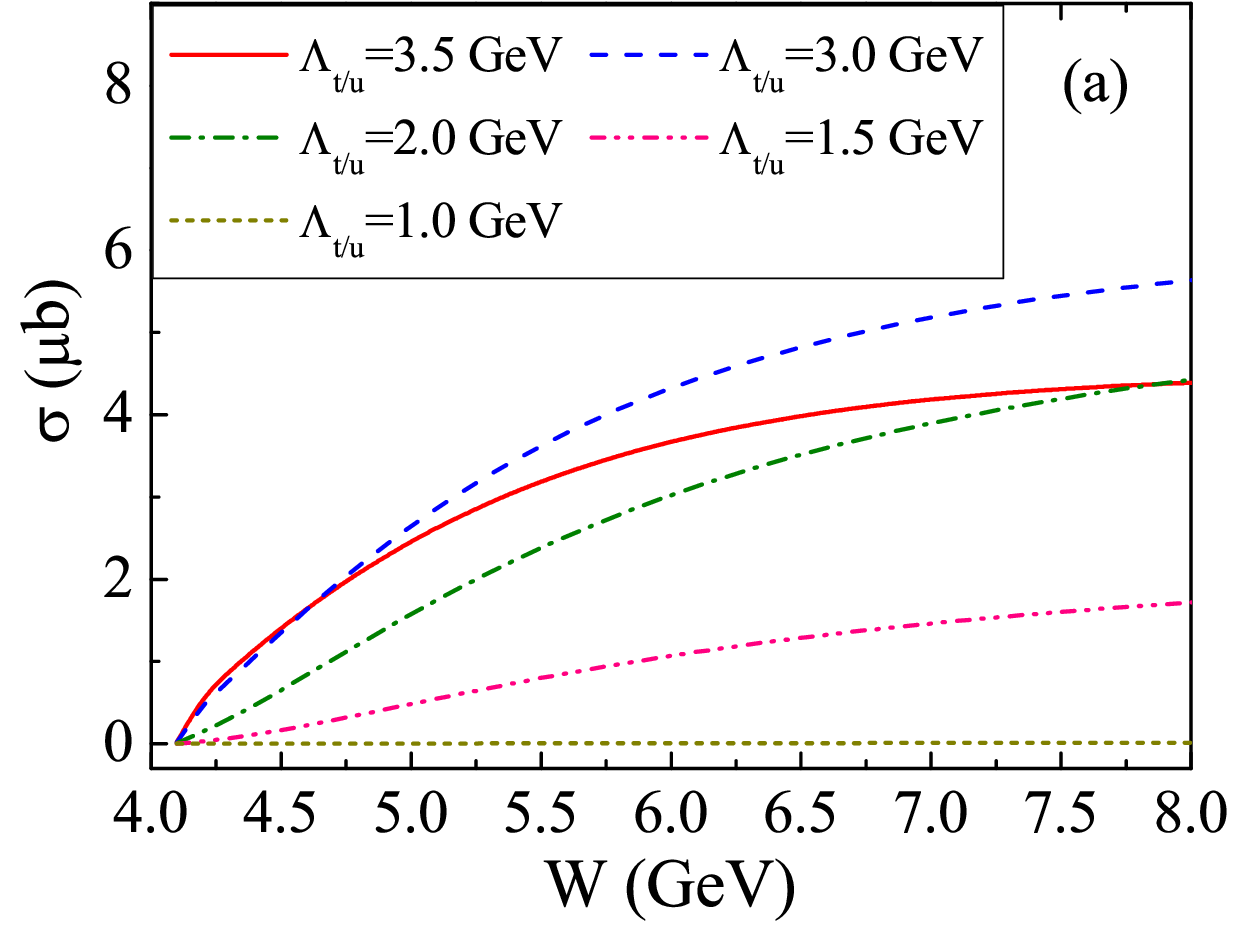} %
\includegraphics[scale=0.35]{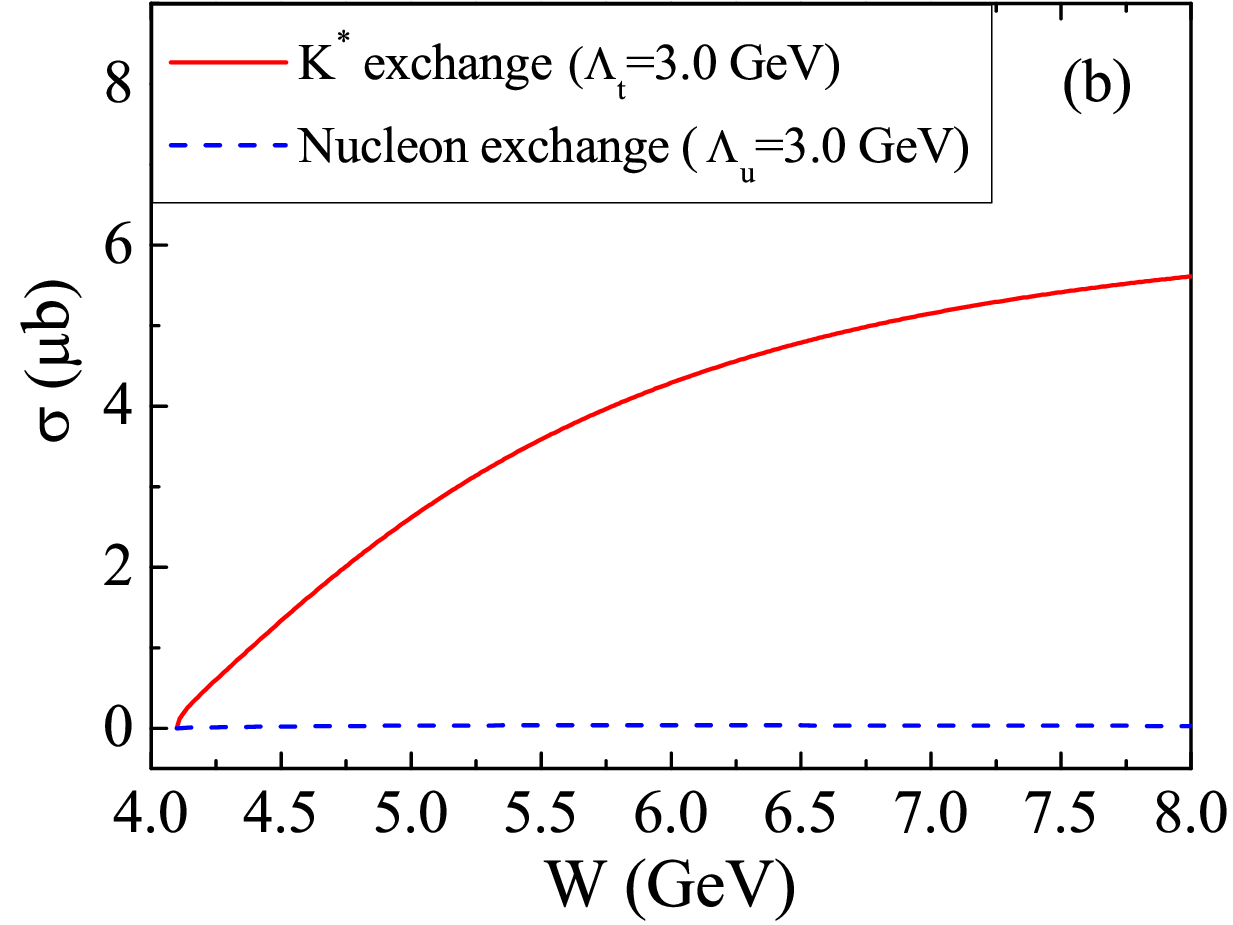}
\end{center}
\caption{(Color online) (a): The cross section for the $K^{-}p\rightarrow
\protect\eta _{c}\Lambda $ reaction from the $K^{\ast }$ and nucleon
exchange contributions with the different values of cutoff parameters $%
\Lambda _{t/u}$. (b) stand for the comparison of background cross section
from $t$-channel with $K^{\ast }$ exchange and $u$-channel with nucleon
exchange.}
\label{Fig:backcut}
\end{figure}

From fig. 2 one notice that the cross section from the signal contributions
is not sensitive to the values of cutoff $\Lambda _{s}$, especially at the
center of mass energy $W\simeq 4.2$ GeV, which is due to the fact that the
form factor for $\Lambda _{c\bar{c}}^{\ast }(4209)$ is close to 1 with the
invariant mass $W$ around $4.2$ GeV in the present calculation. Thus we
constrain the cutoff to be $\Lambda _{s}=2.0$ GeV as used in ref \cite{bc12}%
, which will be used as an input in the following calculations.

Fig. 3(a) present the variation of cross section from the background
contributions for $K^{-}p\rightarrow \eta _{c}\Lambda $ reaction with five
typical $\Lambda _{t/u}$ values (namely, $\Lambda _{t}=\Lambda
_{u}=1.0,1.5,2.0,3.0,3.5$ GeV). It is seen that the cross section of
background reach up to its maximum value when taking $\Lambda _{t/u}=3.0$
GeV, while the cross section of background began to get lower when taking $%
\Lambda _{t/u}=3.5$ GeV. In the spirit of seeking a larger limit for the
cross section of the background, we take $\Lambda _{t/u}=3.0$ GeV in our
estimation\footnote{%
It should be noted that the $K^{-}p\rightarrow \eta _{c}\Lambda $ reaction
through $K^{\ast }$ and nucleon exchange is an OZI forbidden process, which
is similar with the process of $\bar{p}p\rightarrow J/\psi \pi ^{0}$ with
nucleon pole exchange. In our previous work \cite{xy2015}, we have concluded
that the total cross section of $\bar{p}p\rightarrow J/\psi \pi ^{0}$ by
exchanging the nucleon pole are consistent with the E760 and E835 data by
taking cut-off parameter $\Lambda _{N}=1.9$ and $3.0$ GeV, respectively.
Thus it is reasonable to constrain the value of $\Lambda _{t/u}$ to be $3.0$
GeV in the present work.}, which may ensure a reliable estimation on whether
the signal of $\Lambda _{c\bar{c}}^{\ast }(4209)$ can be distinguished from
the background. Moreover, fig. 3(b) show that the $t$-channel with $K^{\ast
} $ exchange play dominant role in the background production and the
contributions from $u$-channel with nucleon exchange is very small.
\begin{figure}[tbp]
\begin{center}
\includegraphics[scale=0.4]{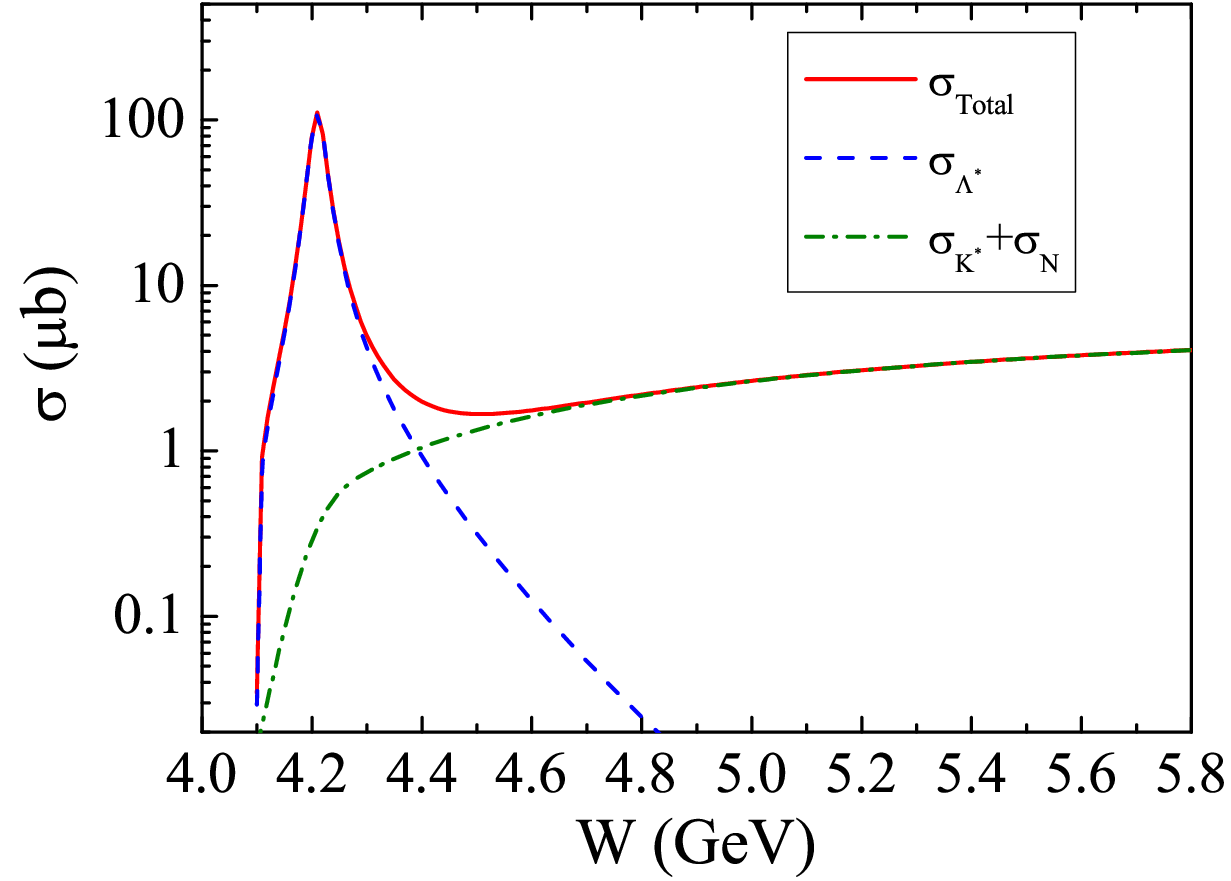}
\end{center}
\caption{(Color online) The total cross section for the $K^{-}p\rightarrow
\protect\eta _{c}\Lambda $ process. Here, the $\protect\sigma _{Total}$ is
the total cross section, while $\protect\sigma _{\Lambda ^{\ast }}$, $%
\protect\sigma _{K^{\ast }}$ and $\protect\sigma _{N}$ stand for the cross
section come from the $\Lambda _{c\bar{c}}^{\ast }(4209)$, $K^{\ast }$ and
nucleon exchange, respectively.}
\label{Fig:total}
\end{figure}

Fig. 4 present the total cross section for $K^{-}p\rightarrow \eta
_{c}\Lambda $ reaction including both signal and background contributions by
taking $\Lambda _{s}=2.0$ GeV and $\Lambda _{t}=\Lambda _{u}=3.0$ GeV. One
notice that the line shape of the total cross section goes up very rapidly
and has a peak around $W\simeq 4.2$ GeV. In this energy region, the cross
section from the intermediate $\Lambda _{c\bar{c}}^{\ast }(4209)$ exchange
is two order larger than that from the background related to the $K^{\ast }$
and nucleon exchange, which indicate that the signal can be clearly
distinguished from the background. Accordingly, we conclude that $W\simeq
4.2 $ GeV is the best energy window of searching for the $\Lambda _{c\bar{c}%
}^{\ast }(4209)$ baryon via $K^{-}p$ collision.

Considering the final state $\eta _{c}$ is a resonance and usually detected
by its decay modes (such as $K\bar{K}\pi $ \textit{et al.}), it is important
to give an estimation about the ratio of $\sigma (K^{-}p\rightarrow \eta
_{c}\Lambda \rightarrow K\bar{K}\pi \Lambda )/\sigma (K^{-}p\rightarrow K%
\bar{K}\pi \Lambda )$. As presented in fig. 4, the cross section of $\Lambda
_{c\bar{c}}^{\ast }(4209)$ is on the order of 110 $\mu b$ around center of
mass energy $W\simeq 4.2$ GeV. Using the branching ratio $BR(\eta
_{c}\rightarrow K\bar{K}\pi )=$ $7.3\%$, we obtain the cross section $\sigma
(K^{-}p\rightarrow \eta _{c}\Lambda \rightarrow K\bar{K}\pi \Lambda )\simeq 8
$ $\mu b$ at $W\simeq 4.2$ GeV. For the total cross section of the $%
K^{-}p\rightarrow K\bar{K}\pi \Lambda $ process, several experiment datas
are available \cite{data}, which are listed in table II.
\begin{table}[tbp]
\caption{The cross section for the $K^{-}p\rightarrow K\bar{K}\protect\pi %
\Lambda $ process at different beam momentum. }%
\begin{tabular}{ccc}
\hline\hline
Reaction & P$_{lab}$ $($GeV$/c)$ & Cross section ($\mu b$) \\ \hline
$K^{-}p\rightarrow \pi ^{0}K^{+}K^{-}\Lambda $ & 2.58 & 4.5 \\
& 4.20 & 84 \\ \hline
$K^{-}p\rightarrow \pi ^{0}2K^{0}\Lambda $ & 6.0 & 39 \\
& 6.5 & 39.3 \\ \hline
$K^{-}p\rightarrow \pi ^{-}K^{+}K^{0}\Lambda $ & 4.25 & 25.2 \\
& 14.3 & 13 \\ \hline\hline
\end{tabular}%
\end{table}

In $K^{-}p$ system, center of mass energy $W\simeq 4.2$ GeV corresponds to $K
$ beam momentum of $P_{lab}\simeq 8.8$ GeV. Thus we estimate that the total
cross section of $K^{-}p\rightarrow K\bar{K}\pi \Lambda $ is on the order of
$40\sim 80$ $\mu b$ at $P_{lab}\simeq 8.8$ GeV. Accordingly, we get the
ratio at $W\simeq 4.2$ GeV as follows,%
\begin{equation*}
\frac{\sigma (K^{-}p\rightarrow \eta _{c}\Lambda \rightarrow K\bar{K}\pi
\Lambda )}{\sigma (K^{-}p\rightarrow K\bar{K}\pi \Lambda )}\simeq 10\sim 20\%%
\text{.}
\end{equation*}

As mentioned above, the J-PARC facility is an idea platform searching for
the predicted $\Lambda _{c\bar{c}}^{\ast }(4209)$ baryon via $K^{-}p$
scattering \cite{jparc}. Assuming the acceptance of $K^{-}p\rightarrow K\bar{%
K}\pi \Lambda $ reaction at J-PARC is about 50\% \cite{jparc}, one can
expect about $4\times 10^{5}\sim 8\times 10^{5}$ events per day for the
production of $K\bar{K}\pi \Lambda $ at $W\simeq 4.2$ GeV, in which about 8$%
\times 10^{4}$ events per day are related to the $\Lambda _{c\bar{c}}^{\ast
}(4209)$.
\begin{figure*}[tbph]
\begin{minipage}{1\textwidth}
\includegraphics[scale=0.38]{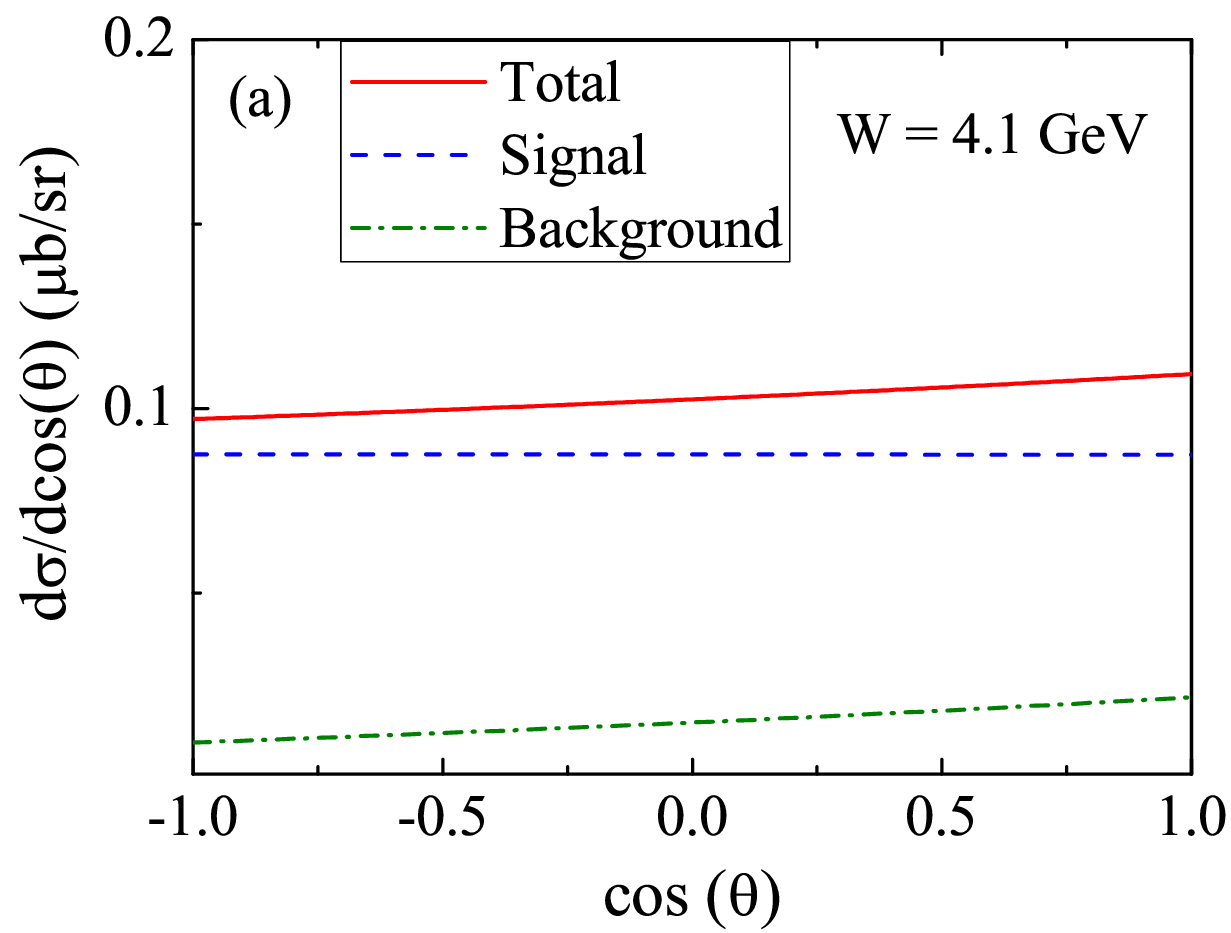}
\includegraphics[scale=0.38]{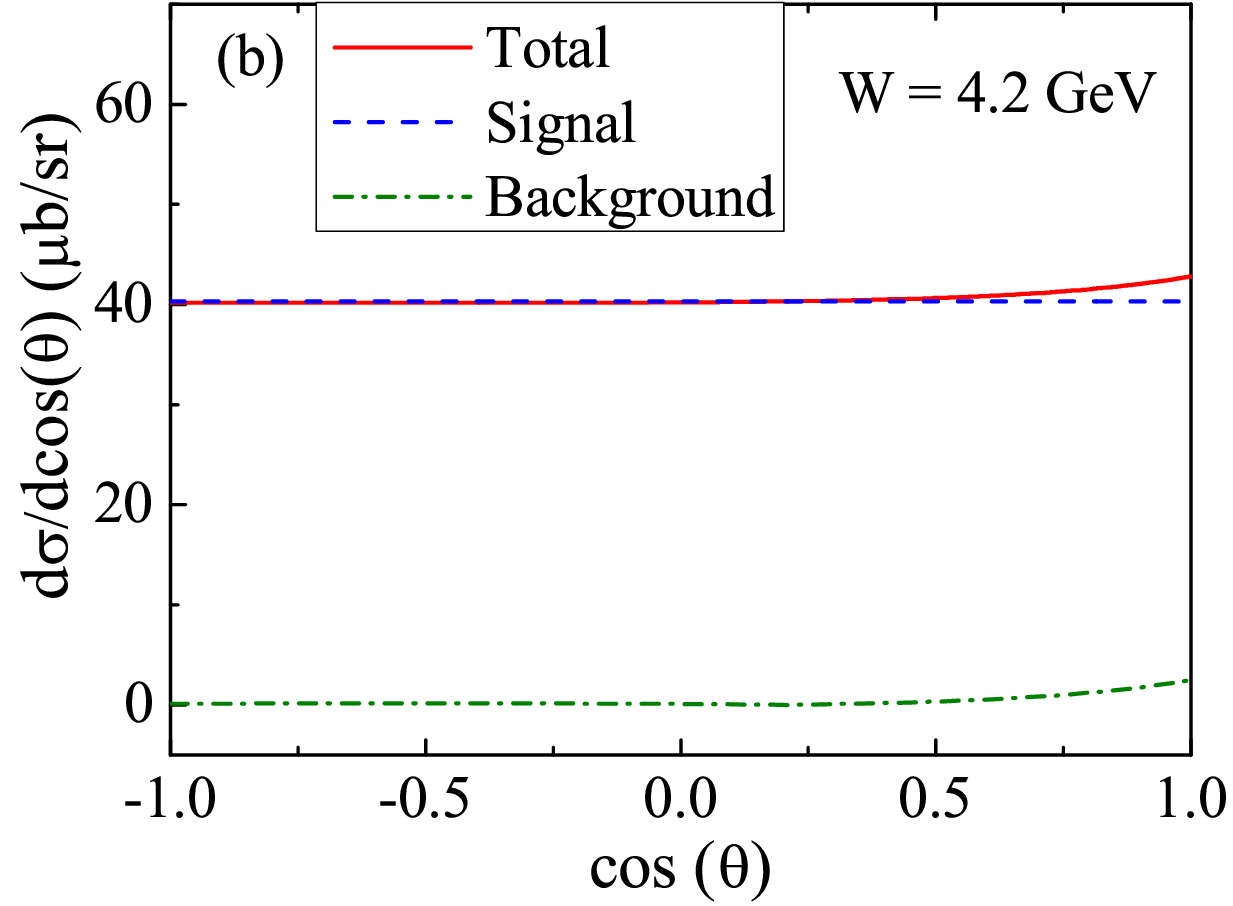}
\includegraphics[scale=0.38]{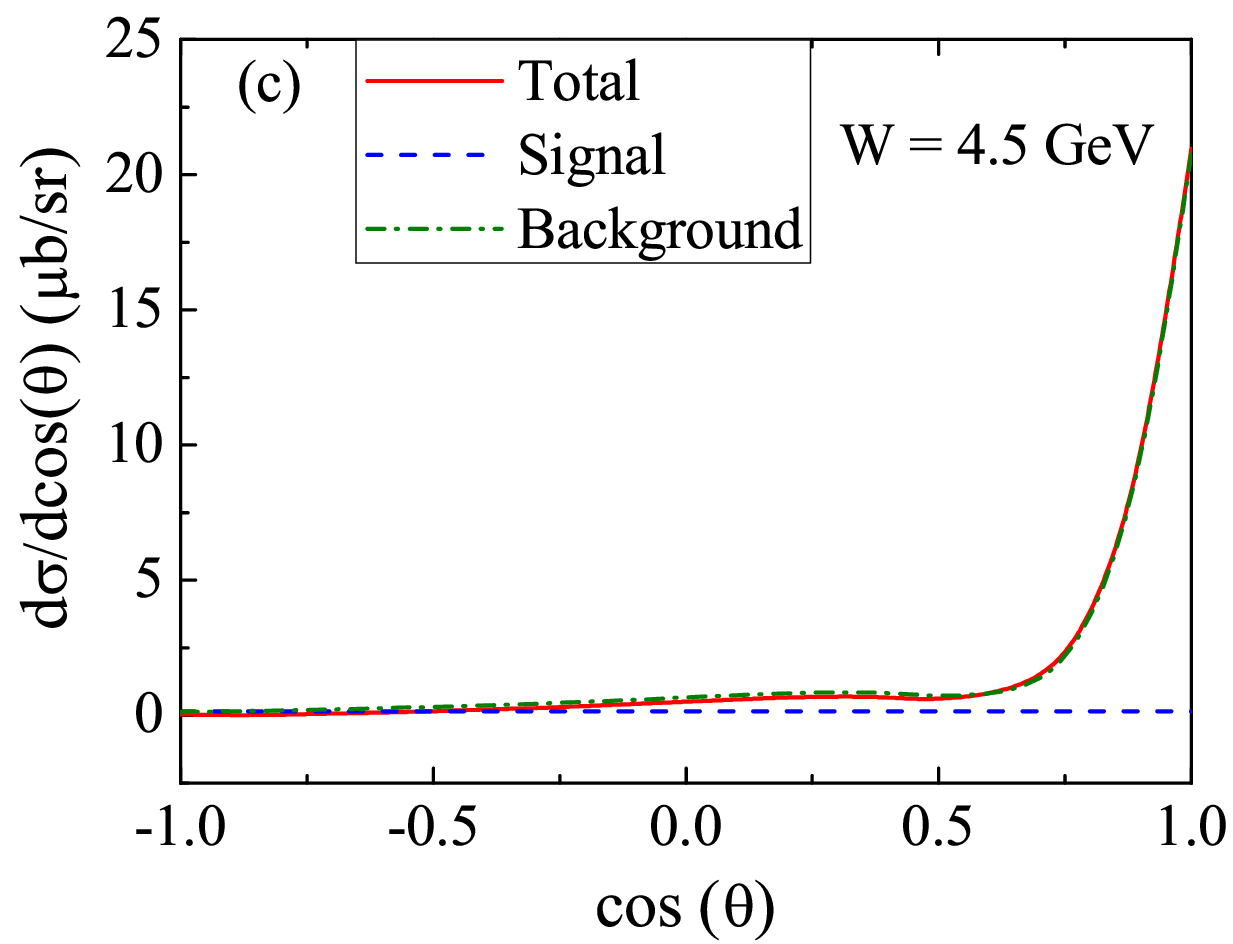}
\includegraphics[scale=0.38]{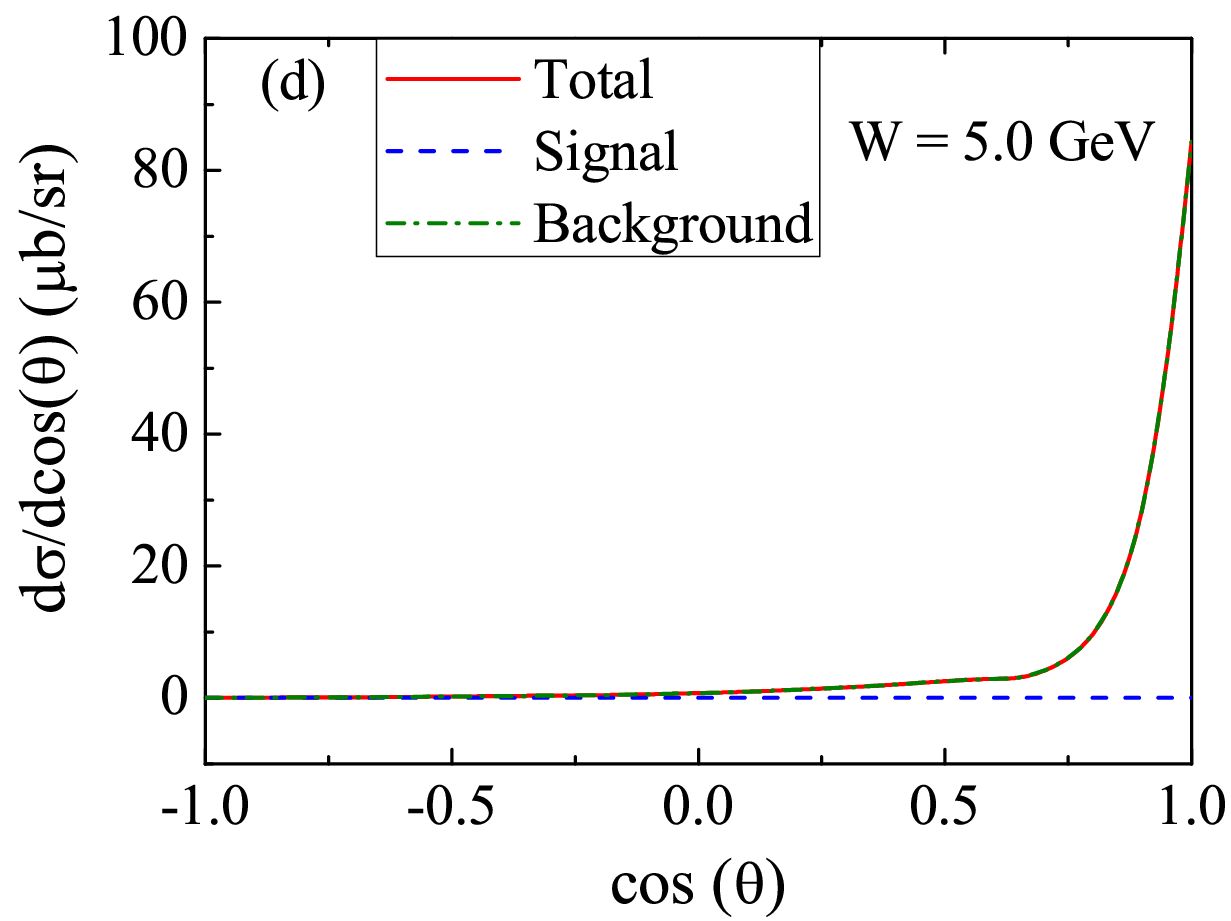}
\caption{(Color online) The differential cross sections for the process of $K^{-}p\rightarrow \eta _{c}\Lambda $ at different
center of mass energy $W=4.1$, $4.2$, $4.5$ and $5$ GeV, where the "Total"
denotes the differential cross section including both signal and background
contributions.}
\end{minipage}
\end{figure*}

We also present the differential cross section for $K^{-}p\rightarrow \eta
_{c}\Lambda $ reaction including both signal and background contributions at
different energy, as shown in fig. 5. It is seen that the differential cross
section of background are sensitive to the $\theta $ angle and gives a
considerable contribution at forward angles. Besides, one notice that the
shapes of the $s$-channel $\Lambda _{c\bar{c}}^{\ast }(4209)$ and the
background are much different, which can be tested in future experiment.

\section{Summary}

Within the frame of the effective Lagrangian approach and isobar model, the
production of superheavy $\Lambda _{c\bar{c}}^{\ast }(4209)$ baryon in the $%
K^{-}p\rightarrow \eta _{c}\Lambda $ process via $s$-channel is
investigated. Moreover, the $t$-channel with $K^{\ast }$ and $u$-channel
with nucleon exchange are also considered, which are regarded as the
background for the $\Lambda _{c\bar{c}}^{\ast }(4209)$ production in the $%
K^{-}p\rightarrow \eta _{c}\Lambda $ reaction.

The numerical results indicate:

\begin{itemize}
\item[(i)] An obvious peak appear in the total cross section of $%
K^{-}p\rightarrow \eta _{c}\Lambda $ reaction near the threshold of $\Lambda
_{c\bar{c}}^{\ast }(4209)$ when the $s$-channel with intermediate $\Lambda
_{c\bar{c}}^{\ast }(4209)$ are included. The cross section from the
intermediate $\Lambda _{c\bar{c}}^{\ast }(4209)$ exchange is two order
larger than that from the background around center of mass energy $W\simeq
4.2$ GeV, which means it is feasible to searching for the predicted $\Lambda
_{c\bar{c}}^{\ast }(4209)$ baryon via the $K^{-}p\rightarrow \eta
_{c}\Lambda $ reaction.

\item[(ii)] The $t$-channel with $K^{\ast }$ exchange play dominant role in
the background production and the contributions from $u$-channel with
nucleon exchange is very small. Besides, the differential cross section of
background are sensitive to the $\theta $ angle and gives a considerable
contribution at forward angles.

\item[(iii)] In the best energy window of $W\simeq 4.2$ GeV, the signal can
be clearly distinguished from the background, while there will be a sizable
number of events related to the $\Lambda _{c\bar{c}}^{\ast }(4209)$ produced
at J-PARC facility.
\end{itemize}

As a final note, the near future experiment at COMPASS@CERN \cite{cern} will
also be enough to check our predictions. Thus we suggest that this
experiment be carried out at the above experimental facilities, which not
only helps in testing the existence of the $\Lambda _{c\bar{c}}^{\ast
}(4209) $ baryon but also provides important information for better
understanding of the production mechanism of the exotic baryons.

\section{Acknowledgments}

This project is partially supported by the National Basic Research Program
(973 Program Grant No. 2014CB845406), the National Natural Science
Foundation of China (Grants No. 11175220) and the One Hundred Person Project
of Chinese Academy of Science (Grant No. Y101020BR0).

\end{document}